\documentclass[mathleft, fleqn, final
]{an}
\usepackage{graphicx}
\usepackage{times}
\usepackage{amsmath}
\usepackage{amssymb}
\begin{document}

% The following seven commands are intended for editorial usage and should be ignored by
% the author(s).
\Pagespan{000}{}% Document's page range. 
% If second parameter is left empty, the last page is computed automatically.
\Yearpublication{2016}%
\Yearsubmission{2015}%
\Month{11}%   
\Volume{XXX}%  
\Issue{XX}% 
% \DOI{This.is/not.aDOI}% 

\title{Gravitational Lensing Size Scales for Quasars}

%\title{Astronomische Nachrichten -- \\
 %       instructions for authors using \LaTeXe\ markup\,\thanks{Data
%from STELLA}}

\author{G. Chartas\inst{1}\fnmsep\thanks{Corresponding author:
  \email{chartasg@cofc.edu}\newline},
%Example 
%for footnote, note the usage of the \texttt{fnmsep}
%command as separator between institute number and footnote mark} 
C.~Rhea\inst{1}, C.~Kochanek\inst{2}, X.~Dai\inst{3}, C.~Morgan\inst{4}, J.~Blackburne\inst{2}, B. Chen\inst{3}, A.~Mosquera\inst{2,4}, \and C.~MacLeod\inst{4,5}
}
\titlerunning{Gravitational Lensing Size Scales for Quasars}
\authorrunning{Chartas et. al.}
\institute{
Department of Physics and Astronomy, College of Charleston, Charleston, SC 29424, USA
\and
Department of Astronomy, The Ohio State University, Columbus, OH 43210, USA
\and
Homer L. Dodge Department of Physics and Astronomy, The University of Oklahoma, Norman, OK 73019, USA
\and
Physics Department, United States Naval Academy, Annapolis, MD 21403
\and
Institute for Astronomy, University of Edinburgh, Edinburgh EH9 3HJ, U.K.}

\received{1 September 2015}
\accepted{16 September 2015}
%\publonline{later}

%\keywords{Editorial notes -- instruction for authors}


\abstract{%
We review results from our monitoring observations of several lensed quasars performed in the optical, UV, and X-ray bands. Modeling of the multi-wavelength light curves provides constraints on the extent of the optical, UV, and X-ray emission regions. One of the important results of our analysis is that the optical sizes as inferred from the microlensing analysis are significantly larger than those predicted by the theoretical-thin-disk estimate. In a few cases we also constrain the slope of the size-wavelength relation. Our size constraints of the soft and hard X-ray emission regions of quasars indicate that in some objects of our sample the hard X-ray emission region is more compact than the soft and in others the soft emission region is smaller. This difference may be the result of the relative strengths of the disk-reflected (harder and extended) versus corona-direct (softer and compact) components in the quasars of our sample. Finally, we present the analysis of several strong microlensing events where we detect an evolution of the relativistic Fe line profile as the magnification caustic traverses the accretion disk. These caustic crossings are used to provide constraints on the innermost stable circular orbit (ISCO) radius and the accretion disk inclination angle of the black hole in quasar RX~J1131$-$1231.}
%independent estimates of the size of Fe K emission region and the spin of the black hole in quasar RX~J1131-1231.}

\maketitle

\section{Introduction}

Due to gravitational lensing the event horizon of a black hole casts a shadow that for a distant observer has a size of about 10$r_{\rm g}$  (e.g., Falcke et al. 2000). The predicted angular sizes of these black-hole shadows for SgrA* and M87 based on their measured black-hole masses are 
$\sim$ 50$\mu$as and $\sim$ 35$\mu$as, respectively.
Direct imaging has been employed to image the accretion disk of SgrA* with submm Very Long Baseline Interferometry (Doeleman et al. 2008, 2011).
%Specifically, measurements using a three station VLBI array at $\lambda = 1.3 mm$  constrain the size of SgrA* to 3.7$^{+1.6}_{-1.0} R_{\rm Sch}$ (Doeleman et al. 2008).  Future improvements of the Event Horizon Telescope will extend VLBI to shorter wavelengths of 0.8mm and 0.65mm where
%the array resolution is expected to approach 10$\mu$as (Doeleman et al. 2011).  

Direct imaging of the environments of quasars using submm VLBI is not possible due to their large distances, however, we currently rely on indirect methods to infer the structure near black holes. Indirect mapping methods include light travel time arguments, reverberation mapping of the broad line region (Blandford \& McKee 1982; Peterson 1993, Netzer \& Peterson 1997), reverberation mapping of the Fe K$\alpha$ emission region (Young \& Reynolds 2000), reverberation between the X-ray and optical/UV continua (Shappee et al. 2014; Edelson et al. 2015),
and microlensing of the continuum and line emission regions (e.g., Grieger et al. 1988 and 1991; Schneider, Ehlers \& Falco 1992; Gould \& Gaudi 1997; 
Agol \& Krolik 1999; Yonehara et al. 1999; Mineshige \& Yonehara 1999; Chartas et al. 2002, 2009, 2012; Popovic et al. 2003, 2006; Blackburne et al. 2006, 2011, 2014, 2015; Pooley et al. 2006, 2007; Kochanek et al. 2004, 2007; Jovanovic et al. 2008; Morgan et al. 2006, 2008a, 2008b, 2010, 2012; Dai et al. 2010;
Mosquera et al. 2011, 2013; Chen et al. 2011, 2012; MacLeod et al. 2015).

Macrolensing of a quasar into multiple images may occur in cases where there is near alignment of the 
observer, an intervening galaxy and the background quasar. The surface mass density of the lensing galaxy needs to be above a critical value before 
multiple images are produced (strong lensing). Typical separations of images produced in macrolensing range between 0.1 -- 20 arcsec.

Microlensing is the bending of light produced by the individual stars in the lensing galaxy.
Microlensing variability occurs when the complex pattern of caustics produced by stars in the lens 
moves across the source plane. 

The characteristic scale of these patterns is the Einstein radius of
\begin{equation}
{R_{\rm E} =  D_{\rm OS} \left[  {{4G \langle M \rangle }\over{c^{2}}}  { {D_{\rm LS}}\over{D_{\rm OL}D_{\rm OS}   }}  \right ]^{1/2}} %= 2.5 \times 10^{16}\left( {{ \langle M \rangle }\over{0.3M_{\odot}} }\right)^{1/2}~{\rm cm}
\end{equation}

\noindent
where $\langle M \rangle$ is the mean mass of the lensing stars, D is the angular diameter distance, and the subscripts L, S, and O refer to the lens, source, and observer, respectively.

Microlensing will affect the images differently resulting in uncorrelated variability between images.
Emission regions with sizes significantly smaller than the projected Einstein radius of the stars will be strongly affected by microlensing 
whereas emission regions with sizes significantly larger will not be affected.
The light curves of compact sources are therefore expected to show enhanced uncorrelated variability compared to the light-curves of 
more extended emission regions.

Simulations of light-curves of images of gravitationally lensed quasars from caustic crossings are fit to observed light-curves and provide constraints 
on the size of the emission regions, the microlens mass scale, and the mass fraction of the local surface density of the lens galaxy. 
The microlensing analysis includes the creation of many random realizations of the star fields near each image and the generation of magnification maps.
This technique was first developed by Kochanek (2004; Kochanek et al. 2007) and has successfully been applied to several gravitational lens systems. 
Recent improvements of microlensing simulations that allow for movement of the stars between epochs also provide constraints on the 
inclination of the accretion disk and the direction of motion of the caustics (e.g., Poindexter \& Kochanek 2010).

The majority of the optical and UV continuum emission of AGN is thought to originate from the accretion disk.
The microlensing method therefore applied to the optical and UV light curves of images of lensed quasars places constraints on the sizes of the accretion disks at their respective rest-frame wavelengths. 
The majority of the X-ray continuum of quasars is dominated by non-thermal radiation of the X-ray corona with a smaller 
contribution from the disk. The accretion disk components is often referred to as the reflected component.
Measurements using the microlensing method to fit the X-ray light curves of the images of lensed quasars 
place constraints on the size of the X-ray hot corona. 

A promising technique for measuring the ISCO  and spin of AGN 
relies on  modeling the Fe K${\alpha}$ fluorescence lines originating from the inner parts of the disk (e.g., Fabian et al. 1989; Laor 1991; Reynolds \& Nowak 2003).  Modeling of the relativistic iron line can only be applied to a few relatively nearby Seyferts where the  line is detectable at a high signal-to-noise level. The Fe K${\alpha}$ line is most Seyferts is typically very weak and constraining the spin and accretion disk parameters of Seyferts  requires considerable observing time on {\it XMM-Newton} and {\it Chandra}.

Here we provide for the first time a new technique based on microlensing that provides a robust upper limit on the size of the 
inner radius of the X-ray emitting region of an accretion disk. Whereas relativistically broadened  Fe~K${\alpha}$ lines detected in the spectra of unlensed AGN are produced from emission originating from a large range of azimuthal angles and radii, the microlensed Fe~K${\alpha}$ lines in the spectra of lensed AGN are produced from a relatively smaller region on the disk that is magnified as a microlensing caustic crosses the disk. 
Simulations of microlensing caustics crossing the accretion disk show significant redshifts and blueshifts of the fluorescent Fe~K${\alpha}$ line (e.g., Popovic et al. 2006). The evolution of the energy and shape of the Fe~K${\alpha}$ line during a caustic crossing depends on the ISCO, spin, inclination angle of the disk and caustic angle. The extreme shifts are produced when the microlensing caustic is near the ISCO of the black hole.
 Measurements of the distribution of the energy shifts of the Fe~K${\alpha}$ line due to microlensing therefore
 provides a powerful and robust method for estimating the ISCO, spin and inclination angle of the disk.
 In \S 2 we present a review of our multi-wavelength monitoring campaign of gravitationally lensed quasars and summarise 
our major results. In \S 3 we present an estimate of the ISCO of  RX~J1131$-$1231 based on the measured distribution of energy shifts of
the  Fe~K${\alpha}$ line caused by microlensing. Finally in \S 4 we present a summary of results obtained from our monitoring campaign 
of lensed quasars.
Throughout this paper we adopt a flat $\Lambda$ cosmology with 
$H_{0}$~=~67~km~s$^{-1}$~Mpc$^{-1}$,  $\Omega_{\rm \Lambda}$ = 0.69, and  $\Omega_{\rm M}$ = 0.31, based on the Planck 2015 results 
(Planck Collaboration et al. 2015).

\begin{table}
% \centering%%%
\caption{Gravitationally Lensed Quasars Monitored}
\label{t1}
\begin{tabular}{lcccc}\hline\hline
Object               & z${_s}$ & z${_l}$        & M$_{BH}$ &  Line  \\ 
                          &             &                                  & (10$^{8}$ M$_{\odot}$) &   \\ 
 \hline                          
Q~J0158$-$4325         & 1.294    &0.317                 &  1.6               & (MgII, a)  \\
HE~0435$-$1223         & 1.689    &0.454                   &   5.0             &  (CIV, a)       \\
SDSS0924+0219     & 1.524    &0.390                   &   2.8               & (MgII, b)   \\
SDSS1004+4112     & 1.734    &0.680                 &   0.4              &  (MgII, b) \\
QSO~1104$-$1805       & 2.319    &0.729                  &   5.9               & (H$\beta$, c) \\
PG~1115+080           & 1.720     &0.310                      &    4.6            &    ( H$\beta$, c) \\
RX~J1131$-$1231        & 0.658  &0.295                 &   0.6              &   (H$\beta$, a)  \\
Q 2237+030             & 1.690     &0.040                   &   12.0               & (H$\beta$, c)\\
 \hline
\hline
\end{tabular}
\\
Notes: \\
{a}-Peng et al. 2006, {b}-Morgan et al. 2006,  {c}-Assef et al. 2011
\end{table}

\section{Multiwavelength Monitoring of Lensed Quasars}
We  are performing multiwavelength monitoring of several quasars listed in Table~\ref{t1},% Table 1,
with the main scientific goal of measuring the emission structure near the black holes in the optical, UV, and X-ray bands in order to test accretion disk models. 
The X-ray monitoring observations were performed with the {\sl Chandra X-ray Observatory}. The Optical (B, R and I band) observations were made with the SMARTS Consortium 1.3m telescope in Chile.
The UV observations were performed with the {\it Hubble Space Telescope}. Here we highlight several of the most recent results published by our microlensing team.

In Blackburne et al. 2015 we analyze the lightcurves of $z$ = 2.32 quasar HE~1104$-$1805 using dynamical microlensing magnification patterns.
HE~1104$-$1805 has been observed at a variety of wavelengths ranging from the mid-infrared to X-ray for nearly 20 years. 
Dynamical microlensing analysis constrains the half-light radius of the accretion disk
to $\log$(r$_{1/2}$/cm) = 16.0$^{+0.30}_{-0.56}$ , the half-light radius of the X-ray emission region
to have an upper limit of $\log$(r$_{1/2}$/cm) = 15.33 (95\% confidence), a low inclination angle is preferred statistically, 
the mean mass of the stars in the lensing galaxy ($<$M$>$) ranges between 0.1 and 0.4~M$_{\odot}$
and the slope of the size-wavelength relation $r_{1/2}$~$\propto$~$\lambda^{\xi}$, is $\xi$ = 1.0$^{+0.30}_{-0.56}$. 
The majority of the observed continuum X-ray emission is found to originate within $\sim$ 30$r_{\rm g }$, assuming a black hole
estimate of $M_{\rm BH}$ = 5.9 $\times$ 10$^{8}$ M$_{\odot}$  based on the width of the H$\beta$ line (Assef et al. 2011).
Based on this black hole mass estimate the gravitational radius of HE~1104$-$1805 is $r_{\rm g }$ = 8.7 $\times$ 10$^{13}$ cm.

In MacLeod et al. 2015 we analyze the light-curves of $z$ = 1.524 quasar SDSS~0924$+$0219 using static microlensing magnification patterns.
SDSS~J0924$+$0219 has been observed at a variety of wavelengths ranging from the near-infrared to X-ray.
Our microlensing analysis in this system constrains the soft-X-ray, UV, and optical half-light radii to be 
2.5$^{+10}_{-2}$ $\times$10$^{14}$  cm, 8$^{+24}_{-7}$ $\times$ 10$^{14}$ cm, and $\sim$ $5^{+5.}_{-2.5}$ $\times$ 10$^{15}$ cm, respectively. Assuming the MgII based  black-hole estimate of $M_{\rm BH}$ = 2.8 $\times$ 10$^{8}$ M$_{\odot}$   the majority of the soft X-ray emission of SDSS~0924$+$0219 originates within $\sim$ 30 $r_{\rm g }$. The gravitational radius of SDSS~0924$+$0219 is $r_{\rm g }$ = 4.12 $\times$ 10$^{13}$ cm.

In Dai et al. 2010 and Chartas et al. 2009 we analyze the light-curves of $z$ = 0.658 quasar RX~J1131$-$1231.
We find the X-ray and optical half-light radii to be 2.3 $\times$ 10$^{14}$~cm and 1.3 $\times$ 10$^{15}$~cm, respectively.
These sizes correspond to $\sim$ 26$r_{\rm g }$ and $\sim$ 147$r_{\rm g }$, respectively.
%The gravitational radius of the black hole of RX~J1131 is estimated (based on H${\beta}$) to be $\sim$ 2. $\times$ 10^${13}$~cm.

An important result found in all microlensing studies is that optical sizes of quasar accretion disks as inferred from the microlensing analysis are significantly larger than those predicted by thin-disk theory. Specifically, measurements of the radius of the accretion disk at 2500$\rm \AA$ rest-frame 
indicate that the sizes obtained from microlensing measurements are 2--3 times larger than the values predicted by thin-disk theory
(e.g., Morgan et al. 2010, Mosquera et al. 2013).

%as a function of black hole mass 

\begin{figure}
\begin{center}
\includegraphics[width=80mm,height=40mm]{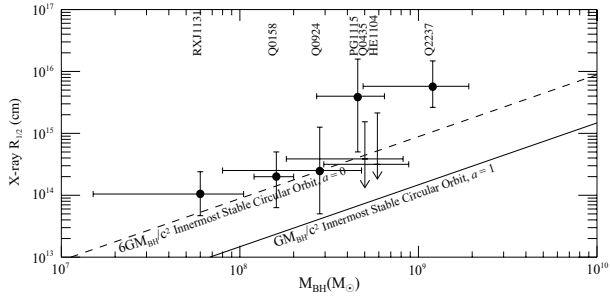}
\end{center}
\caption{X-ray half-light radii of quasars as determined from our microlensing analysis versus their black hole masses.
}
\label{label1}
\end{figure}

In Figure 1 we present the X-ray half-light radii of quasars from recent microlensing studies of lensed systems observed as part of our monitoring program (MacLeod et al. 2015, Blackburne et al. 2014, 2015,  Mosquera et al. 2013, Morgan et al. 2008, 2012, Dai et al. 2010, and Chartas et al. 2009).
Included in Figure 1 are the uncertainties of the black-hole mass estimates and uncertainties in the size estimates.
%In most cases the inclination angles of the AGN disks have been assumed to satisfy cosi = 0.5. 
The X-ray sizes of the quasars in our sample are found to be close to the sizes of their innermost stable circular orbits.
Assuming that most of the X-ray emission in the band detected originates from the hot X-ray corona, these results indicate that the corona is very compact and not extended over a large portion of the accretion disk.

\section{Estimating the Inner Most Stable Circular Orbit Using Microlensing}

RX~J1131$-$1231 has been monitored 38 times over a period of 10 years with the {\sl Chandra X-ray Observatory}.
As reported in Chartas el al. 2012,  redshifted and blueshifted Fe Ka lines have been detected in the spectra of the lensed images.

\begin{figure}
\begin{center}
\includegraphics[width=80mm,height=100mm]{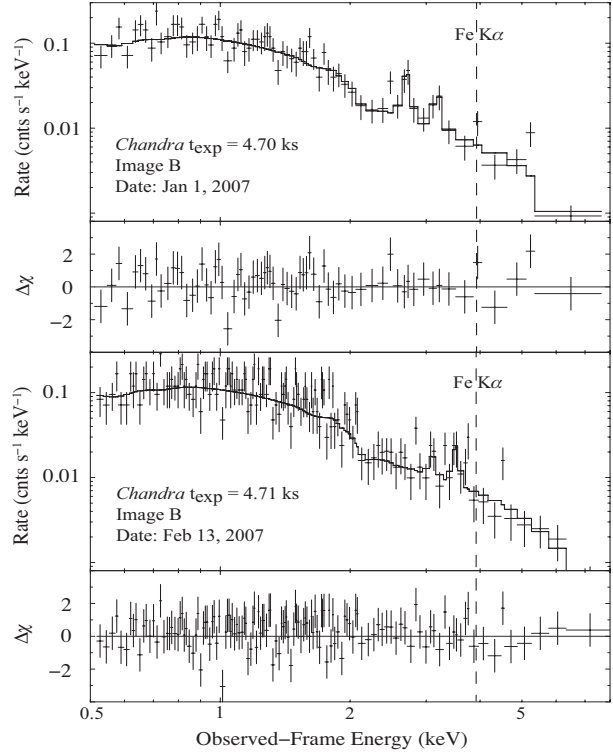}
\end{center}
\caption{Evolution of the Fe K$\alpha$ line possibly caused by the motion of a magnification caustic as it moves away from the center of the black hole.}
\label{label1}
\end{figure}

\begin{figure}
\begin{center}
\includegraphics[width=80mm,height=100mm]{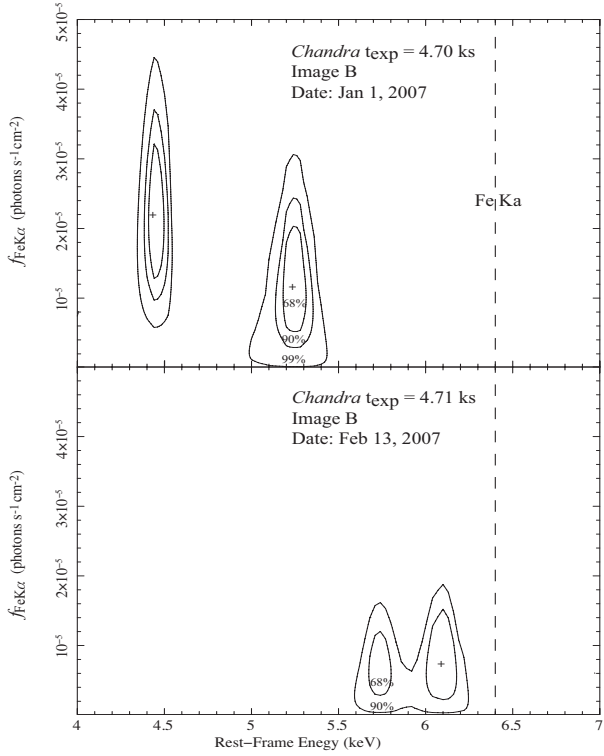}
\end{center}
\caption{ The contours represent 68\%, 90\%, and 99\% $\chi^{2}$  confidence intervals of the flux normalizations of the detected Fe~K$\alpha$ line in the spectra (Figure 2)  of image B for the Jan 1, and Feb 13 observations of RX~J1131$-$1231.
}
\label{label1}
\end{figure}

\begin{figure}
\begin{center}
\includegraphics[width=80mm,height=60mm]{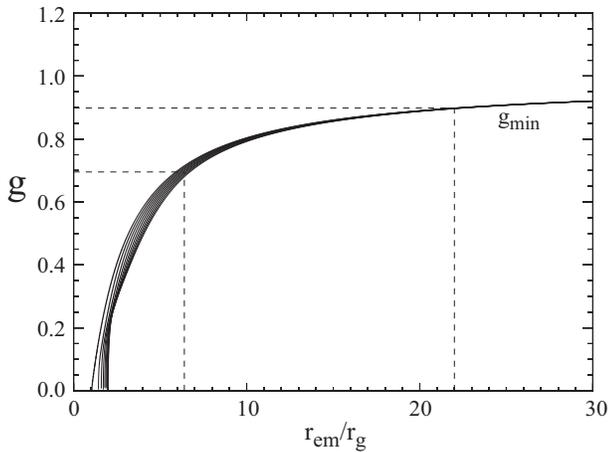}
\end{center}
\caption{Extremal shifts of the observed Fe K$\alpha$ line energy for spin values ranging between 0.1 and 0.998 in increments of 0.1.
Horizontal dashed lines represent the observed values of the energy shifts $g$ = $E_{\rm obs}$/$E_{\rm rest}$ of the most redshifted  Fe  K$\alpha$ line components of the two epochs presented in Figure 2. The observed shifts of the energy of the Fe K$\alpha$ line can be interpreted as the motion of a magnification caustic over a distance of $\sim$ 16r$_{g}$ between the two epochs.}
\label{label1}
\end{figure}

\begin{figure}
\begin{center}
\includegraphics[width=80mm,height=120mm]{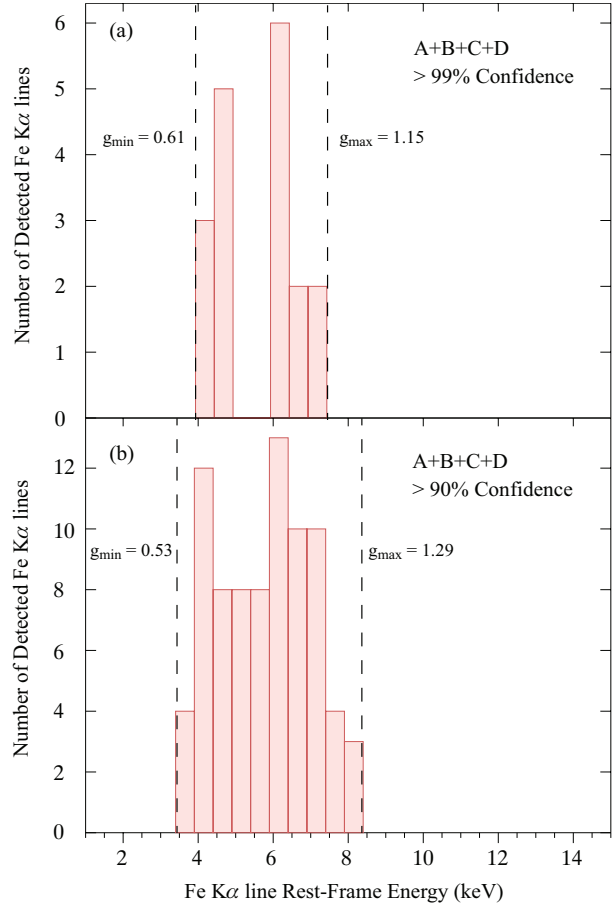}
\end{center}
\caption{Distribution of the energy-shifts of the Fe~K$\alpha$ line in all images from all 38 epochs of RX~J1131$-$1231. 
Only cases where the iron line is detected at $\geq$ 99\% (panel a) and at $\geq$ 90\% (panel b)  are shown. The vertical lines represent the extreme cut-offs of the 
distribution. These cut-offs provide upper limits of the ISCO and inclination angle of the black hole.}
\label{label1}
\end{figure}

\begin{figure}
\begin{center}
\includegraphics[width=80mm,height=60mm]{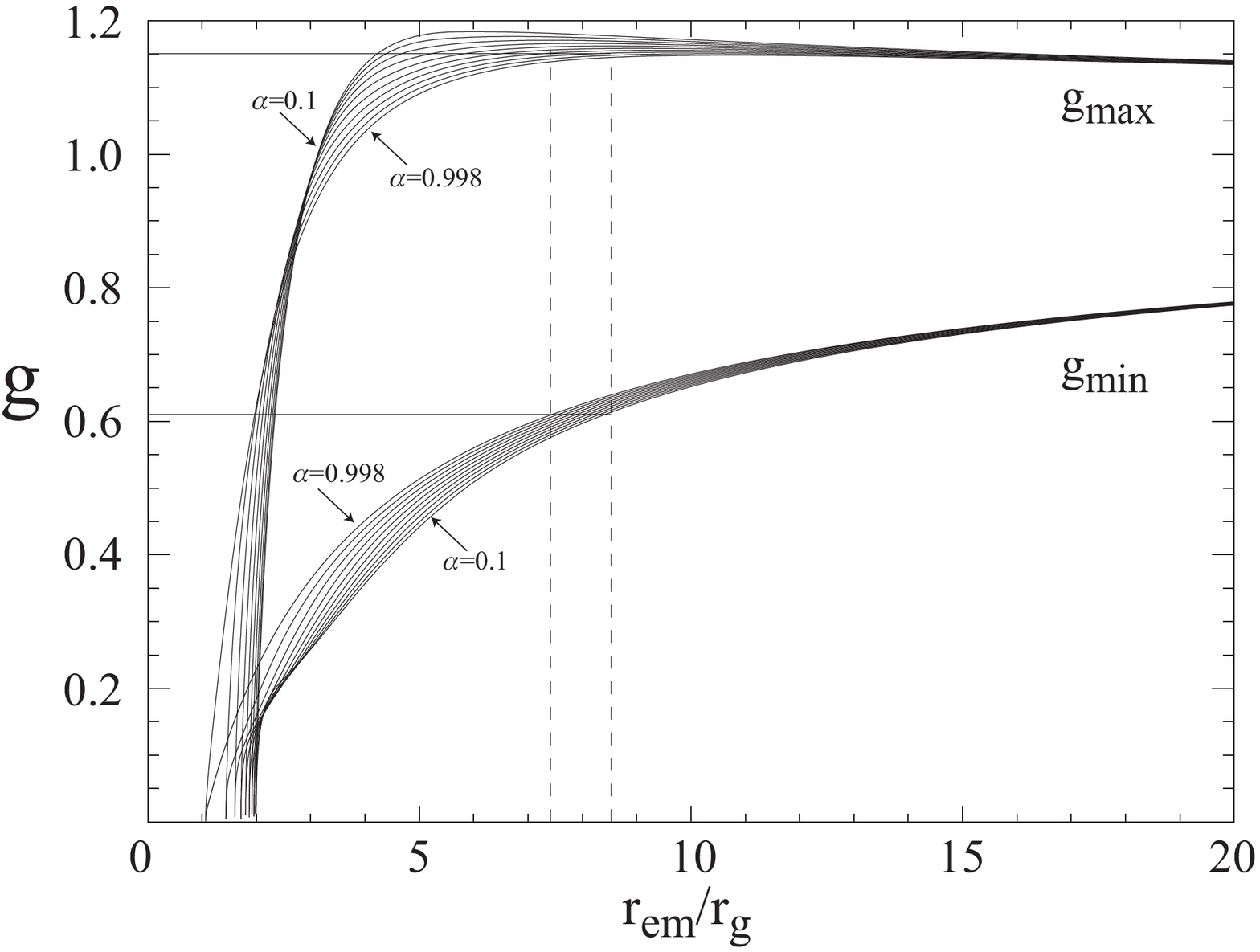}
\end{center}
\caption{ Extremal shifts of the observed Fe K$\alpha$ line energy for spin values ranging between 0.1 and 0.998 in increments of 0.1.
Horizontal solid lines represent the observed values of the normalized energy shifts $g$ = $E_{\rm obs}$/$E_{\rm rest}$ of the most redshifted and blueshifted Fe  K$\alpha$ lines
from all 38 epochs and all images of RX~J1131$-$1231. The  extreme $g$ values are for Fe  K$\alpha$ lines detected at $\geq$ 99\%.
The inner radius of the accretion disk is constrained to be $r  <  8.5r_{g}$.}
\label{label1}
\end{figure}

In Figure 2 we show the evolution of the red and blue components of the Fe K$\alpha$ line possibly caused by the motion of a magnification caustic as it moves away from the center of the black hole.  
We interpret the shift of the Fe K$\alpha$  line as resulting from general relativistic and special relativistic Doppler effects.
As shown in Figure 3, the two redshifted iron lines in the Jan 1, 2007 observation are each detected at the $\geq$ 99\% confidence and 
the  iron lines in the Feb 13, 2007 observation are each marginally detected at the $\geq$ 90\% confidence level (Figure 3).

The extreme values of the normalized energy shifts $g$ = $E_{\rm obs}$/$E_{\rm rest}$ of the two lines can be used to roughly constrain the radius of the emitting material as shown in Figure 4. 
The generalized Doppler shifts $g$ were calculated for the case of a spinning black hole (Kerr 1963) using the formalism described 
in Muller \& Camenzind 2004 and Karas \& Sochora 2010.

For these calculations we assumed the radial and toroidal velocity components  $v^{(r)}$ and v$^{(\theta)}$  of the radiating plasma of the accretion disk in the  Zero Angular Momentum Observer frame to be relatively small.
For a given inclination angle $i$, the angle between the direction of the orbital velocity $v^{(\phi)}$ of the plasma and our line of sight 
is assumed to range between a minimum and maximum value that will depend on the inclination angle $i$, and the 
angle between the caustic direction of motion and the projection of our line-of-sight onto the disk plane.
We assume that the most redshifted line component, as shown in Figure 2, is produced by Fe~K$\alpha$ emission originating close to the ISCO.
%We assume that these angular extremes correspond to the two peaks of the iron line detected in the spectra shown in Figure 3. 
For $r$ $\geq$ $r_{\rm ms}$ a Keplerian profile is assumed (equation 15 of Muller \& Camenzind 2004), whereas for $r$ $\leq$ $r_{\rm ms}$ constant specific angular momentum is assumed.

The two redshifted iron lines detected in the Jan 1, 2007 observation correspond to generalized Doppler shifts  
of $g_{\rm min}$ $\sim$ 0.7 and $g_{\rm max}$ $\sim$ 0.82  and the two lines  detected in the Feb 13, 2007 observation correspond to $g_{\rm min}$ $\sim$ 0.9 and $g_{\rm max}$ $\sim$ 0.96.
The detected energies of the iron lines and their evolution is consistent with a microlensing caustic moving 
by a distance of about 15$r_{\rm g}$  between the two epochs that are separated by 44 days (observed frame).

We can independently check this estimate by estimating the distance a caustic in this system is expected to travel in 44 days.
Due to the combined motions of the observer, lens, source, and stars in the lens galaxy, the source moves relative to the magnification patterns at an effective velocity of  v$_{\rm eff}$ $\sim$ 700 km~s$^{-1}$ (see Mosquera et al. 2011).  
Assuming a black-hole mass of $\sim$ 6 $\times$  10$^7$ M$_{\odot}$ (i.e., Sluse et al. 2012) the gravitational radius of RX~J1131$-$1231 is
$r_{\rm g}$~=~3.6 ~$\times$ ~10$^{13}$~cm. 
The distance ${\Delta}d_{\rm c}$ travelled by a caustic in ${\Delta}t_{\rm obs}$ = 44 days is
${\Delta}d_{\rm c}$  = v$_{\rm eff}$ ${\Delta}t_{\rm obs}$/(1+$z$) $\sim$ 18$r_{\rm g}$. 
This is consistent with the estimated value based on the shift of the Fe~K$\alpha$ lines between the two epochs.

We have performed a systematic spectral analysis of all epochs and all images searching for the presence of the Fe~K$\alpha$ lines in the spectra. 
To determine the significance of the iron lines detected we produced $\chi^{2}$ confidence contours of the flux normalization of the lines as a function of their energy (see Figure 3). 
The spectral models used to fit the X-ray spectra of RX~J1131$-$1231 consist of a power law with Galactic and neutral intrinsic absorption 
and Gaussian lines to model the iron lines. For producing the  $\chi^{2}$ confidence contours the only model parameter that was frozen in the 
spectral fit was the Galactic column density of $N_{\rm H}$ = 3.6 $\times$ 10$^{20}$ cm$^{-2}$ (Dickey \& Lockman 1990).

We find that the iron line is detected in 58 out of the 152 spectra (38 epochs $\times$ 4 images) at $\geq$ 90\% confidence and in 18/152 spectra at $\geq$ 99\% confidence. There is considerable shift of the energy of the iron line due to microlensing. 
%Because of the moderate signal-to-noise ratio (S/N) of the spectra we used the $C-$statistic to fit the spectra and determine the energy shifts and equivalent widths of the Fe~K$\alpha$ line.

In Figure 5 we show the distribution of the energy-shifts of the Fe~K$\alpha$ line in all images.
% and in each image separately.
%We note that images C and D are fainter than the other images and the detection of an Fe~K$\alpha$ line  in the spectra of images C and D is limited by the moderate to low S/N of their spectra.
One important feature of this iron line energy-shift distribution in the significant cut-off of the distribution  at 
rest-frame energies of $\sim$ 3.9~keV and $\sim$ 7.4~keV for iron lines detected at $\geq 99\%$ confidence. 
These cut-offs represent the most extremely 
redshifted and blueshifted Fe~K$\alpha$  lines.
If we interpret the largest energy-shifts to be produced by X-ray emission originating close to the inner most stable circular orbit of the black hole
we can provide upper limits on the size of the ISCO and inclination angle of RX~J1131$-$1231. 

The extreme blueshift of the $g$ distribution is sensitive to the inclination angle of the disk. 
Specifically, the measured generalized Doppler factor $g_{\rm max}$ = 1.15 (99\% confidence)  
constrains the inclination angle to be $\geq$  55$^{\circ}$.
%The measured Doppler factor $g_{\rm max}$ = 1.29 (90\% confidence) 
%constrains the inclination angle to be $\geq$ 65  degrees.

The extreme redshift of the $g$ distribution is sensitive to the ISCO radius. Specifically, the measured generalized  Doppler factor $g_{\rm min}$ = 0.61 (99\% confidence)  constrains 
3.5$r_{\rm g}$  $<$ $r_{\rm ISCO}$  $<$   9$r_{\rm g}$. For the measured generalized  Doppler factor $g_{\rm min}$ = 0.53 (90\% confidence) we find 3$r_{\rm g}$  $<$ $r_{\rm ISCO}$  $<$   7$r_{\rm g}$.

In Figure 6 we plot the $g$ values of the iron line for an accretion disk with an inclination angle 
of $i = 60^{\circ}$ and a caustic crossing angle of $\theta_{c}$ = 20$^{\circ}$.  The observed extreme values of $g_{\rm min}$ = 0.6  and 
$g_{\rm max}$ = 1.15  from all epochs and all images are consistent with an emitting region of 
an $r_{\rm ISCO}$ $\lesssim$ 9$r_{\rm g}$ and an inclination angle of $\sim$ 60$^{\circ}$. 

A more detailed analysis will involve simulating the shifts of the Fe~K$\alpha$ line due to microlensing caustic crossings for a range of inclination angles, caustic crossing angles and black hole spin parameters.
A comparison between the observed generalized Doppler shift $g$  distribution with those simulated will provide more accurate 
constraints on the inclination angle, and  the ISCO. The spin of the black hole can be inferred from the known relation between ISCO radius and spin.  

Additional monitoring observations with ${\it Chandra}$ will provide more representative and complete distributions of the generalized Doppler shift $g$ values of the Fe~K$\alpha$ line in RX~J1131$-$1231 and other lensed quasars. 
The $g$-distribution method provides a new and independent technique of constraining the inclination angle, ISCO radius, and  black hole spin of quasars.

\section{Conclusions}
Multiwavelegth monitoring of lensed quasars provides a powerful technique for constraining the sizes of accretion disks and hot X-ray coronae
of distant quasars. Simultaneous modeling of the optical, UV and X-ray light curves places strong constraints on their relative sizes.
Recent improvements to this method include dynamical microlensing which provides additional constraints on the inclination angle of the disk.
Several important results of our monitoring campaign of lensed quasars are: 

1. Optical sizes of accretion disks at $2500 {\rm \AA}$ inferred from our microlensing analysis are a factor of 2--3 larger than those predicted by thin disk theory. This result  is consistent with recent 
measurements of continuum lags in nearby Seyferts NGC2617 (Shappee et al. 2014) and NGC5548 (Edelson et al. 2015).

2. The size of the UV emission region is found to be a factor of about 10 on average smaller than the optical but larger than the X-ray emission region.

2. The X-ray corona is found to be compact with a half-flight radius  $<$ 30$r_{\rm g}$ (see Figure 1).

4. The scaling between optical sizes at $2500 {\rm \AA}$  and black hole mass is consistent with thin disk theory.

Future improvements include monitoring a larger variety of quasars (for example, BAL quasars and radio loud quasars.)
Several models predict that radio-loud quasars contain truncated disks and measurements of the sizes of radio-loud objects
may address this issue.

We introduced a promising new technique ($g$-distribution method) for measuring the inclination angle, ISCO, and spin of a black hole.
The $g$-distribution method involves measurements of the distribution of the energy shifts of the relativistic iron line emitted from the accretion disk and a comparison of the measured $g$-distribution with microlensing caustic simulations. The method has been applied to RX~J1131$-$1231 and 
initial results indicate that $r_{\rm ISCO}$ $\lesssim$ 9$r_{\rm g}$ and $i$ $\gtrsim$ 60$^{\circ}$.
Further monitoring of RX~J1131$-$1231 and other lensed quasars will provide tighter constraints on the inclination angle, ISCO radius and spin of the black hole of distant quasars.

%Future improvements are to apply the g-distribution method to additional lensed quasars.

%Measure the Doppler shifts distributions of more lensed quasars

%Identify quasars that have large enough image separation that can be resolved with the 
%next genera ration X-ray observatories (ATHENA).

%LSST is expected to increase the number of detected lensed quasars to ~ (0

%Notes: \\
%{}$^{a}$ Peng et al. (2006), {}$^{b}$ Morgan et al. (2006),   {}$^{c}$ Assef et al. (2011)

\acknowledgements
We acknowledge financial support from NASA via the Smithsonian Institution grant G03-141l0B.

%\newpage%%%%%%%%%%%%%%%%%%%%%%%%%%%%%%%%%%%%%%%%%%%%%%%%%%%%%%

\end{document}